\documentclass{article}

\usepackage{PRIMEarxiv}

\usepackage[utf8]{inputenc} 
\usepackage[T1]{fontenc}    
\usepackage{hyperref}       
\usepackage{url}            
\usepackage{booktabs}       
\usepackage{amsfonts}       
\usepackage{nicefrac}       
\usepackage{microtype}      
\usepackage{graphicx}       
\graphicspath{{figures/}}   
\usepackage{amsmath}
\usepackage{array}
\usepackage{multirow}
\usepackage{enumitem}
\usepackage{xcolor}
\usepackage{colortbl}
\usepackage{listings}
\usepackage{float}
\hypersetup{
  colorlinks=true,
  linkcolor=blue!50!black,
  citecolor=blue!50!black,
  urlcolor=blue!50!black
}

\newcommand{\lightrule}{\addlinespace[2pt]\arrayrulecolor{black!25}\specialrule{0.4pt}{0pt}{0pt}\arrayrulecolor{black}\addlinespace[3pt]}

\title{Agentic Self-Healing for Data \& AI Pipelines: \\ An Affordable Vendor-Agnostic Architecture \\ using Open-Source Software}

\author{
  Solomon Eshun \\
  ishango.ai \\
  Accra, Ghana \\
  \And
  Dennis Murage \\
  ishango.ai \\
  Nairobi, Kenya \\
  \And
  Sharleen Muoki \\
  ishango.ai \\
  Nairobi, Kenya \\
  \And
  Chih-Chun Chen \\
  ishango.ai \\
  Berlin, Germany \\
  \AND
  Stephen Adjignon \\
  ishango.ai \\
  Accra, Ghana \\
  \And
  Matteo Staar \\
  Karlsruhe University of Applied Sciences \\
  Karlsruhe, Germany \\
  \And
  Oliver Ang\'elil \\
  ishango.ai \\
  Zurich, Switzerland \\
}

\begin{document}

\maketitle

\begin{abstract}
Modern organizations rely on data, machine learning, and software delivery pipelines to move data, train models, deploy applications, refresh dashboards, and support business-critical decisions. However, these pipelines often fail because of data quality issues, schema changes, upstream source changes, infrastructure problems, orchestration failures, and model workflow issues. Existing ZeroOps, observability, and AI operations platforms can help teams detect incidents, investigate root causes, and in some cases recommend or execute fixes. However, many of these solutions are expensive, vendor-specific, or difficult for smaller teams to adapt across different tools and environments. This paper first compares existing off-the-shelf solutions for AI-assisted pipeline monitoring, root-cause analysis, and automated remediation, including their strengths, limitations, and practical trade-offs. Based on this comparison, we find that the main gap is architectural rather than technological: the required ingredients for self-healing pipelines already exist, but they are fragmented across vendor-specific platforms, observability tools, incident systems, and open-source components. We therefore propose an affordable, vendor-agnostic reference architecture for agentic self-healing pipelines using open-source and low-cost tools. The proposed architecture combines monitoring, pipeline metadata, incident history, deterministic policy checks, AI-assisted diagnosis, approval workflows, and controlled remediation actions to help teams detect, diagnose, repair, verify, and learn from pipeline issues with less manual effort. The goal is to provide a practical reference architecture that can be adapted across data engineering, machine learning operations, and software delivery environments.
\end{abstract}

\keywords{self-healing pipelines \and AI-assisted operations \and data observability \and root-cause analysis \and automated remediation \and agentic AI \and machine learning operations \and software delivery pipelines \and vendor-agnostic architecture \and open-source software \and incident management \and Agentic Recovery and Incident Response}

\section{Introduction}
\label{sec:introduction}

Data, machine learning, and software delivery pipelines have become the connective tissue of modern organizations. A single business dashboard may depend on ingestion jobs that pull from external APIs, transformation layers built with SQL frameworks, feature pipelines that feed model training, batch scoring jobs, and continuous delivery workflows that deploy the services consuming those predictions. Each link in this chain can fail, and in practice each does: schemas drift as upstream teams change source systems, late or malformed data breaks assumptions baked into transformations, infrastructure degrades or runs out of resources, orchestration dependencies deadlock, and model quality decays silently as the world shifts away from the training distribution~\cite{sculley2015,paleyes2022,sambasivan2021}.

The operational cost of these failures is disproportionately borne by people. Interview studies of machine learning practitioners consistently find that sustaining models in production---diagnosing failures, maintaining pipelines, and responding to incidents---is a recurring and significant engineering burden, distinct from building new capability~\cite{shankar2022}. The typical failure workflow remains largely manual: an alert (or worse, a stakeholder) reports a stale dashboard; an engineer reconstructs the failure by traversing logs, orchestrator UIs, and upstream systems; a fix is applied by hand; and the fix is verified by watching the next run. Each step depends on tacit knowledge held by a small number of senior engineers, making the process slow, stressful, and difficult to scale.

The industry response has been a rapidly growing category of AI-assisted operations tooling. AIOps platforms apply machine learning to event correlation and anomaly detection~\cite{dang2019,notaro2021}; data observability products monitor freshness, volume, schema, and distributional health of data assets; and, most recently, large language model (LLM) agents have been embedded into operations platforms to summarize incidents, propose root causes, and in some cases execute remediations~\cite{ahmed2023,chen2024}. Several vendors now market fully ``ZeroOps'' or autonomous operations experiences. These products are genuinely capable, but they present three recurring adoption barriers for small and medium-sized teams. First, \emph{cost}: pricing is typically per-host, per-table, per-user, or by enterprise contract, and grows quickly with estate size. Second, \emph{vendor coupling}: the most autonomous experiences are only available when the entire pipeline estate lives inside a single vendor's ecosystem. Third, \emph{adaptability}: heterogeneous environments---an Airflow instance here, a dbt project there, a homegrown scoring service, a legacy CI server---rarely fit the happy path these platforms assume.

At the same time, the ingredients for building an equivalent capability have become commodities. Open-source monitoring, lineage, data quality, and workflow tooling are mature; LLM inference is inexpensive and available both as hosted APIs and as locally deployable open-weight models; and agent design patterns such as tool-augmented reasoning are well documented~\cite{yao2023,xi2023}. What is missing is not technology but \emph{architecture}: a reference design that shows how to assemble these commodity parts into a trustworthy self-healing loop with appropriate human oversight.

This paper makes two contributions:

\begin{enumerate}[leftmargin=1.5em]
  \item \textbf{A structured comparison of eight off-the-shelf platforms} for AI-assisted pipeline monitoring, root-cause analysis (RCA), and automated remediation---Databricks Genie ZeroOps, Acceldata ADM, Dynatrace Davis AI with Workflows, Datadog Bits AI with Workflow Automation and Data Observability, Monte Carlo, IBM Databand, PagerDuty SRE Agent with Runbook Automation, and ServiceNow Predictive AIOps with Autonomous Workforce and AI Control Tower---evaluated on scope, autonomy, vendor coupling, and cost (Section~\ref{sec:landscape}).
  \item \textbf{An affordable, vendor-agnostic reference architecture} for agentic self-healing pipelines built from open-source and low-cost components, combining telemetry, pipeline metadata, incident history, deterministic policy rules, LLM-based diagnosis, human approval workflows, and guarded remediation actions (Sections~\ref{sec:architecture} and~\ref{sec:lifecycle}).
\end{enumerate}

The architecture is deliberately prescriptive rather than empirical: it distills recurring patterns from the surveyed platforms and from the AIOps and LLM-agent literature into a design that a team of two to four engineers can stand up incrementally, without committing to any single vendor. We discuss limitations, risks, and buy-versus-build guidance in Section~\ref{sec:discussion}.

\section{Background and Related Work}
\label{sec:background}

\subsection{Why pipelines fail}
\label{sec:failures}

Across data engineering, MLOps, and software delivery, pipeline failures cluster into six recurring classes:

\begin{itemize}[leftmargin=1.5em]
  \item \textbf{Data quality issues.} Null spikes, duplicated records, out-of-range values, and distributional anomalies that violate assumptions embedded in downstream logic. Data cascades of this kind are pervasive and compound silently~\cite{sambasivan2021}.
  \item \textbf{Schema changes.} Columns renamed, retyped, dropped, or added by upstream owners, frequently without notice; the absence of enforced data contracts makes these the canonical breaking change~\cite{jones2023}.
  \item \textbf{Upstream source changes.} API version bumps, altered export cadences, moved file locations, revoked credentials, or semantic changes in how a source system populates a field.
  \item \textbf{Infrastructure problems.} Out-of-memory kills, disk exhaustion, spot-instance preemption, network partitions, expiring certificates, and quota limits.
  \item \textbf{Orchestration failures.} Dependency deadlocks, misconfigured schedules, stuck sensors, backfill collisions, and retry storms in tools such as Airflow~\cite{airflow2025}.
  \item \textbf{Model workflow issues.} Training divergence, feature skew between offline and online paths, prediction drift, and gradual performance decay~\cite{sculley2015,breck2017}.
\end{itemize}

These classes differ in where they surface (data plane, control plane, or model quality plane) but share a diagnostic structure: the observed symptom is usually several hops downstream of the root cause, which is why lineage metadata is central to any effective diagnosis~\cite{openlineage2025}.

\subsection{The self-healing loop}
\label{sec:loop}

We use \emph{self-healing} to mean a closed operational loop with eight stages: \emph{detect} (a monitor or test raises a signal), \emph{triage} (deduplicate, classify, and prioritize), \emph{diagnose} (identify the root cause), \emph{plan} (select a candidate remediation), \emph{approve} (a policy gate decides whether a human must confirm), \emph{remediate} (execute the action), \emph{verify} (confirm the system has recovered), and \emph{learn} (record the episode so future incidents resolve faster). Classical site reliability engineering executes this loop with humans at every stage, supported by runbooks and automation for the mechanical parts~\cite{beyer2016}. The question this paper addresses is how much of the loop can be delegated to software, and under what guardrails, without importing unacceptable risk. In practice this means routing known failures to deterministic rules, and reserving LLM agents for incidents where diagnosis requires contextual reasoning.

\subsection{AIOps and LLM-based incident management}

AIOps research has historically focused on the left half of the loop: anomaly detection, event correlation, and failure prediction over metrics and logs~\cite{dang2019}. Notaro et al.\ survey more than a decade of such methods and note that automated \emph{remediation} remains the least developed stage~\cite{notaro2021}. LLMs have recently shifted this frontier. Ahmed et al.\ show that fine-tuned language models can recommend plausible root causes and mitigation steps for cloud incidents at scale~\cite{ahmed2023}, and Chen et al.\ demonstrate an LLM-based RCA assistant deployed over real production incidents, using retrieval over incident history to ground its diagnoses~\cite{chen2024}. In parallel, the agent literature has converged on a pattern---interleaved reasoning and tool use~\cite{yao2023}---that maps naturally onto operations work: an agent that can query metrics, read logs, inspect lineage, and consult runbooks can perform the same investigation a human would, faster and with perfect recall of prior incidents~\cite{xi2023}. Adjacent tool categories apply the same ideas to narrower domains: Sentry provides AI-assisted error monitoring and debugging for application code~\cite{sentry2026}, and LangSmith Engine detects and diagnoses recurring failures in LLM application traces~\cite{langsmith2026}; both are relevant context, though neither targets self-healing for data and ML pipelines. Commercial operations platforms have productized these advances aggressively, which motivates the comparison that follows.

\section{Off-the-Shelf Solutions: A Comparative Analysis}
\label{sec:landscape}

This section surveys eight commercial offerings that span the data observability, application observability, incident response, and IT service management (ITSM) traditions. We selected these platforms because they represent the major approaches currently used for AI-assisted operations: platform-native autonomous operations, data observability, application and infrastructure observability, incident response automation, and enterprise IT service management. The goal is not to provide an exhaustive market survey, but to compare representative systems that expose the architectural trade-offs relevant to self-healing pipelines: detection, root-cause analysis, remediation autonomy, governance, vendor coupling, and cost. We deliberately exclude \emph{implementation frameworks} such as LangGraph, Pydantic AI, Prefect Marvin, and LangSmith from this comparison: they are building blocks for constructing agentic systems rather than off-the-shelf operations platforms, and we return to them as implementation options in Section~\ref{sec:architecture}~\cite{langgraph2026,pydanticai2026,marvin2026,langsmith2026}.

We evaluate each platform on five dimensions: \emph{primary scope} (which failure classes from Section~\ref{sec:failures} it targets), \emph{detection} and \emph{RCA} capability, \emph{remediation autonomy} (from alert-only to closed-loop execution), \emph{vendor coupling} (how much of the estate must live in the vendor's ecosystem to realize the value), and \emph{indicative cost tier}. Assessments are based on vendor documentation and public product descriptions as of mid-2026~\cite{databricks2025,acceldata2025,dynatrace2025,datadog2025,montecarlo2025,ibm2025,pagerduty2025,servicenow2025}; capabilities evolve quickly, so the comparison should be read as a snapshot of category structure rather than a procurement verdict.

\subsection{Platform summaries}

\textbf{Databricks Genie ZeroOps.} Databricks embeds agentic assistance directly into its Data Intelligence Platform: Genie provides conversational access to data and operational context, while Genie ZeroOps---announced in mid-2026 and initially available in preview---is a background agent that autonomously monitors, investigates, and proposes fixes for jobs, pipelines, tables, and ML workloads~\cite{databricks2025}. Because the platform owns the compute, the catalog (Unity Catalog), the lineage graph, and the orchestrator, its agents act with unusually rich context: they can trace failures through lineage, validate candidate fixes against cloned data in an isolated environment, and apply them once a user approves. The corresponding limitation is scope: the autonomy applies to workloads running \emph{on Databricks}. Pipelines that traverse external orchestrators, warehouses, or delivery systems fall outside the healing boundary, and adopting the capability implies adopting the platform.

\textbf{Acceldata ADM.} Acceldata's Agentic Data Management positions a fleet of specialized agents over an enterprise data observability substrate: agents monitor data reliability, diagnose pipeline and infrastructure issues, propose corrective actions, and execute approved playbooks~\cite{acceldata2025}. It is notable for covering both the data plane (quality, freshness, schema) and the compute plane (Spark, warehouse performance, cost), and for supporting multiple underlying stacks. It remains a proprietary enterprise product: the agent layer, the metadata store, and the automation runtime are all vendor-operated, and pricing follows enterprise data observability norms.

\textbf{Dynatrace Davis AI + Workflows.} Dynatrace pairs its causal AI engine (Davis) with an AutomationEngine that triggers workflows from AI-identified problems~\cite{dynatrace2025}. Davis's strength is deterministic causal correlation over a richly instrumented topology model (Smartscape), which yields high-precision root-cause identification for application and infrastructure failures; workflows can then execute remediations such as rollbacks, restarts, or ticket enrichment. The platform is strongest where its OneAgent instrumentation runs---application services, hosts, Kubernetes---and comparatively thin on data-pipeline-native concerns such as table-level quality, dbt models, or training workflows. Consumption-based pricing across hosts, logs, and events is a recurring cost-management theme for adopters.

\textbf{Datadog Bits AI + Workflow Automation + Data Observability.} Datadog combines an LLM assistant (Bits AI) that investigates incidents conversationally, a Workflow Automation product with hundreds of action integrations, and a growing Data Observability offering that adds table freshness, volume, and quality monitoring to its core telemetry platform~\cite{datadog2025}. The breadth is attractive for teams already standardized on Datadog, and the workflow builder makes human-approved remediation practical. Costs, however, accumulate per product module and per host/GB, and the AI features primarily reason over data already inside Datadog---estates with significant telemetry outside the platform see reduced value.

\textbf{Monte Carlo.} Monte Carlo largely defined the data observability category: ML-driven monitors for freshness, volume, schema, and field-level quality, automatic lineage, and incident management with impact analysis~\cite{montecarlo2025}. Its RCA support---correlating an anomaly with recent code, data, or infrastructure changes---is strong within the data warehouse/lakehouse perimeter. Remediation remains largely human-executed: the platform excels at detection, triage, and routing, but does not aim to repair pipelines autonomously. Pricing is enterprise SaaS, typically scoped by monitored tables/users, which smaller teams often find difficult to justify.

\textbf{IBM Databand.} Databand focuses on pipeline observability for data engineering stacks (notably Airflow and Spark): run-level monitoring, SLA tracking, data quality checks, and alerting with lineage context~\cite{ibm2025}. It provides solid detection and diagnostic context for orchestration and data quality failures but stops short of automated remediation, and as part of the IBM portfolio it is typically procured within larger enterprise agreements. Its pipeline-run-centric data model is a good conceptual fit for the incident store we propose in Section~\ref{sec:architecture}.

\textbf{PagerDuty SRE Agent + Runbook Automation.} PagerDuty approaches the problem from incident response: the SRE Agent performs agentic triage and investigation when an incident fires---gathering telemetry from connected monitoring tools, summarizing probable cause, and proposing next steps---while Runbook Automation (formerly Rundeck) provides audited, parameterized execution of operational procedures across arbitrary infrastructure~\cite{pagerduty2025}. This combination is comparatively vendor-neutral: PagerDuty orchestrates over whatever monitoring and infrastructure a team already has. Its blind spot is the data plane---it has no native notion of tables, schemas, or model quality---so data incidents must be surfaced by third-party monitors before PagerDuty adds value. Per-user and per-capability pricing is moderate relative to enterprise observability suites.

\textbf{ServiceNow Predictive AIOps + Autonomous Workforce + AI Control Tower.} ServiceNow offers the most complete closed loop of the surveyed platforms: Predictive AIOps correlates events and detects anomalies against a CMDB topology; AI agents (the ``Autonomous Workforce'') investigate and execute remediations through the platform's workflow engine; and AI Control Tower provides centralized governance---inventory, policy, and audit---over all AI agents operating in the enterprise~\cite{servicenow2025}. The governance layer is genuinely differentiated and anticipates regulatory scrutiny of autonomous operations. The trade-offs are equally clear: the loop runs on ServiceNow's platform primitives (CMDB, ITSM workflows), realizing value requires deep platform adoption, and licensing sits firmly at the enterprise tier. Like PagerDuty, it is infrastructure- and service-centric rather than data-pipeline-native.

\subsection{Synthesis}
\label{sec:synthesis}

\begin{figure}[H]
  \centering
  \includegraphics[width=0.98\textwidth]{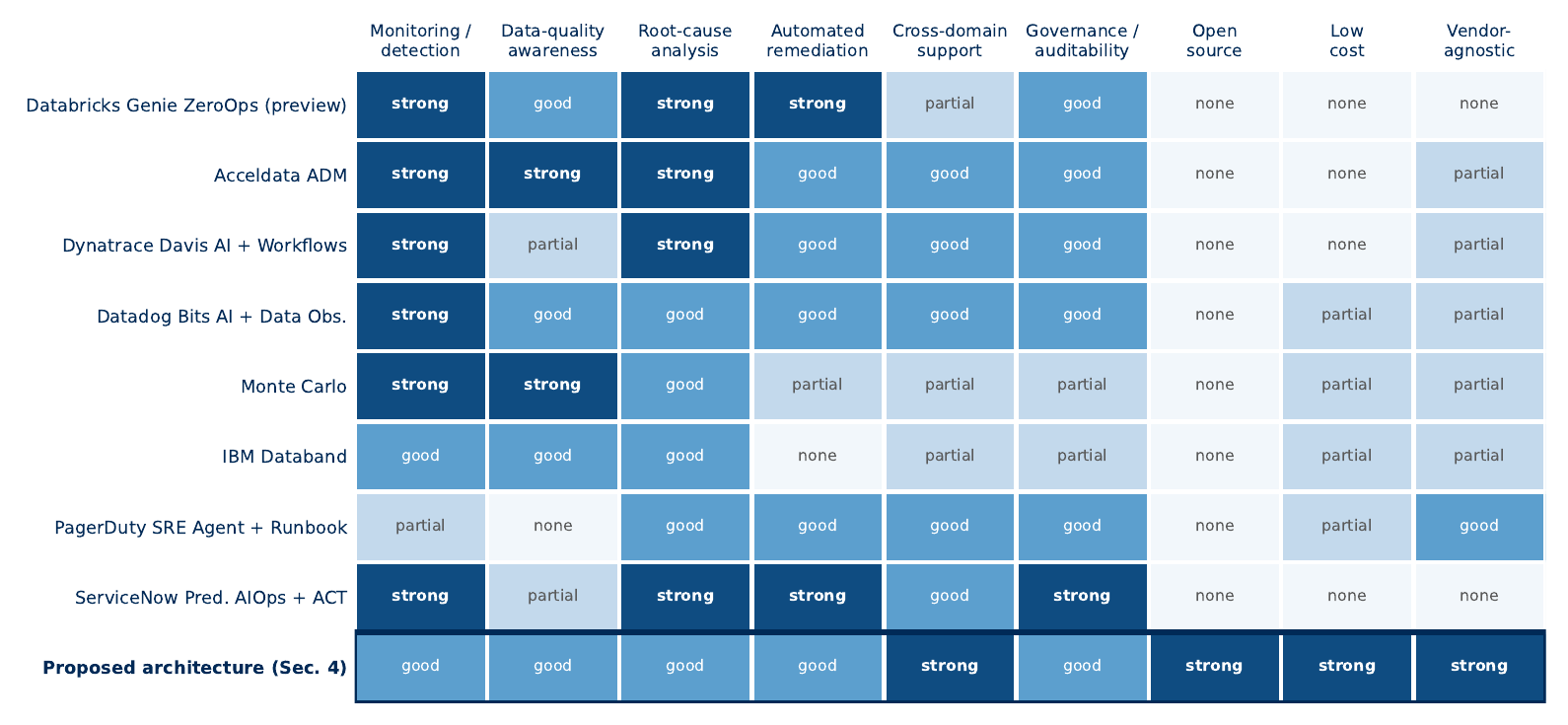}
  \caption{Capability comparison of the eight surveyed solutions, plus the proposed architecture (bold border), across dimensions relevant to agentic self-healing. Cells are scored on an ordinal scale---none, partial, good, and strong---based on publicly available vendor documentation and product descriptions as of mid-2026~\cite{databricks2025,acceldata2025,dynatrace2025,datadog2025,montecarlo2025,ibm2025,pagerduty2025,servicenow2025}. The scoring is qualitative and intended to summarize category-level trade-offs rather than provide a procurement ranking.}
  \label{fig:heatmap}
\end{figure}

\begin{figure}[H]
  \centering
  \includegraphics[width=0.88\textwidth]{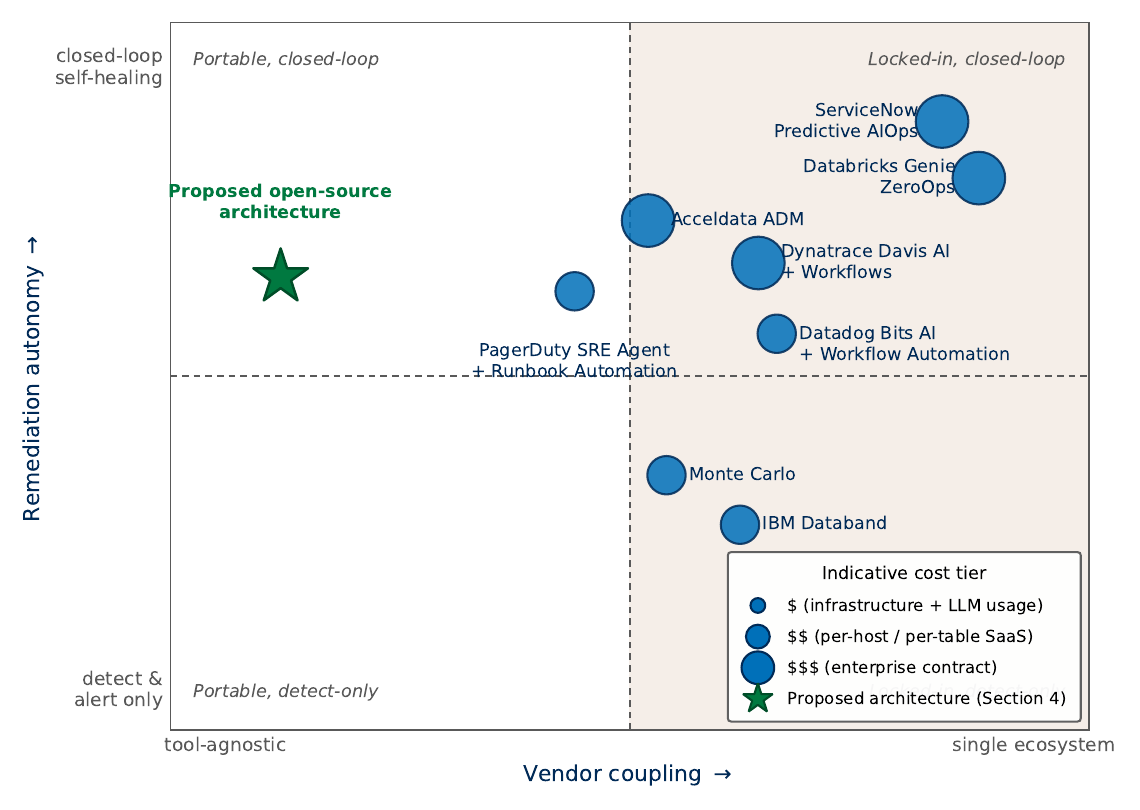}
  \caption{Positioning of the surveyed platforms by vendor coupling (horizontal) and remediation autonomy (vertical); marker size indicates indicative cost tier. Axis placements are the authors' qualitative assessments from vendor documentation, not measured quantities. The most autonomous experiences (upper right) require the deepest ecosystem commitment; the proposed architecture targets the underserved upper-left region.}
  \label{fig:landscape}
\end{figure}

To make the comparison easier to scan, Figure~\ref{fig:heatmap} summarizes the eight platforms---plus the proposed architecture introduced in Section~\ref{sec:architecture}, included here as a reference point---across capability dimensions relevant to self-healing pipeline operations. Scores follow a simple rubric: \emph{strong} indicates the capability is explicitly documented with a described mechanism; \emph{good} indicates partial or inferred support; \emph{partial} indicates the capability is mentioned but narrow in scope; \emph{none} indicates no public evidence of the capability. The scores are qualitative and based on public documentation; they are not intended as a procurement benchmark, but highlight broad architectural patterns across the market. Databricks Genie ZeroOps was still in preview at the time of writing and is assessed from its initial announcement rather than mature documentation, so its scores should be read as preliminary. Figure~\ref{fig:landscape} then positions the eight surveyed platforms on the two axes that matter most for the architecture question: remediation autonomy and vendor coupling.

Three structural observations emerge:

\begin{enumerate}[leftmargin=1.5em]
  \item \textbf{Autonomy correlates with lock-in.} The platforms that close the loop most completely (Databricks, ServiceNow) do so precisely because they own the surrounding context---catalog, orchestrator, CMDB, workflow engine. Autonomy is easy when the healing boundary coincides with the platform boundary; it is the heterogeneous estate that is hard.
  \item \textbf{No single product covers all three pipeline domains.} Data observability products (Monte Carlo, Databand) are weak on remediation and blind to application delivery; application observability products (Dynatrace, Datadog) are thin on table- and model-level semantics; ITSM/incident products (ServiceNow, PagerDuty) depend on others for detection. Teams routinely need two or three subscriptions to cover the failure classes of Section~\ref{sec:failures}.
  \item \textbf{The mechanisms are convergent and reproducible.} Underneath the branding, every platform combines the same five ingredients: telemetry, metadata/topology, an inference layer (statistical, causal, or LLM), a policy/approval gate, and an action runtime. Each ingredient now has a mature open-source counterpart. This is the observation the remainder of the paper builds on.
\end{enumerate}

\section{A Vendor-Agnostic Reference Architecture}
\label{sec:architecture}

Based on the findings from the platform comparison, we propose \textbf{Agentic Recovery and Incident Response}, a vendor-agnostic reference architecture for detecting, diagnosing, repairing, verifying, and learning from failures in data, machine learning, and software delivery pipelines. This architecture is not a fixed software product or a mandatory tool stack; it is a modular architecture that defines the responsibilities, interfaces, and safety controls required for agentic self-healing workflows. Throughout this section we name concrete tools, but they are illustrative rather than prescriptive: the architecture is defined by interfaces and responsibilities, not by a fixed technology stack, and a practical implementation should select the smallest viable set of tools that fits the organization's existing environment.

\subsection{Design principles}

The architecture is shaped by six principles, chosen to invert the trade-offs identified in Section~\ref{sec:synthesis}:

\begin{itemize}[leftmargin=1.5em]
  \item \textbf{P1 --- Vendor agnosticism through open interfaces.} Every integration point uses an open standard or open-source tool: OpenTelemetry for traces and logs, Prometheus exposition for metrics, OpenLineage for lineage events~\cite{openlineage2025,prometheus2025}. The estate being healed (orchestrators, warehouses, CI systems) is treated as replaceable---and so is every component of the architecture itself.
  \item \textbf{P2 --- The LLM is a swappable commodity.} All agent calls go through a model gateway so that hosted APIs, model routing providers, and locally served open-weight models---the large language models (LLMs) themselves, e.g., GPT, Claude, or Llama-family models---are interchangeable per task, controlling cost, latency, and data residency.
  \item \textbf{P3 --- Deterministic before generative.} Not every incident requires LLM reasoning. Known low-risk failures first pass through a deterministic policy layer that applies rules, thresholds, playbooks, and risk classifications; LLM-based agents are reserved for incidents whose diagnosis requires reasoning across logs, lineage, incident history, or ambiguous evidence. This reduces cost, improves predictability, and limits hallucination risk.
  \item \textbf{P4 --- Guarded autonomy, not full autonomy.} Agents may only execute actions drawn from an explicit, version-controlled allowlist, and every action class carries a risk tier that determines whether human approval is required. This follows the SRE principle that automation must be at least as auditable as the human procedure it replaces~\cite{beyer2016}.
  \item \textbf{P5 --- Memory is the moat.} The system's compounding value comes from its incident history: every episode---symptom, diagnosis, action, approval, outcome---is stored and retrieved to ground future diagnoses, mirroring the retrieval-grounded RCA shown effective in production settings~\cite{chen2024}.
  \item \textbf{P6 --- Incremental adoption.} Each layer of the architecture is independently useful. A team can deploy monitoring alone, add diagnosis later, and enable actuation last, once trust is established.
\end{itemize}

\subsection{Architecture overview}

\begin{figure}[H]
  \centering
  \includegraphics[width=0.98\textwidth]{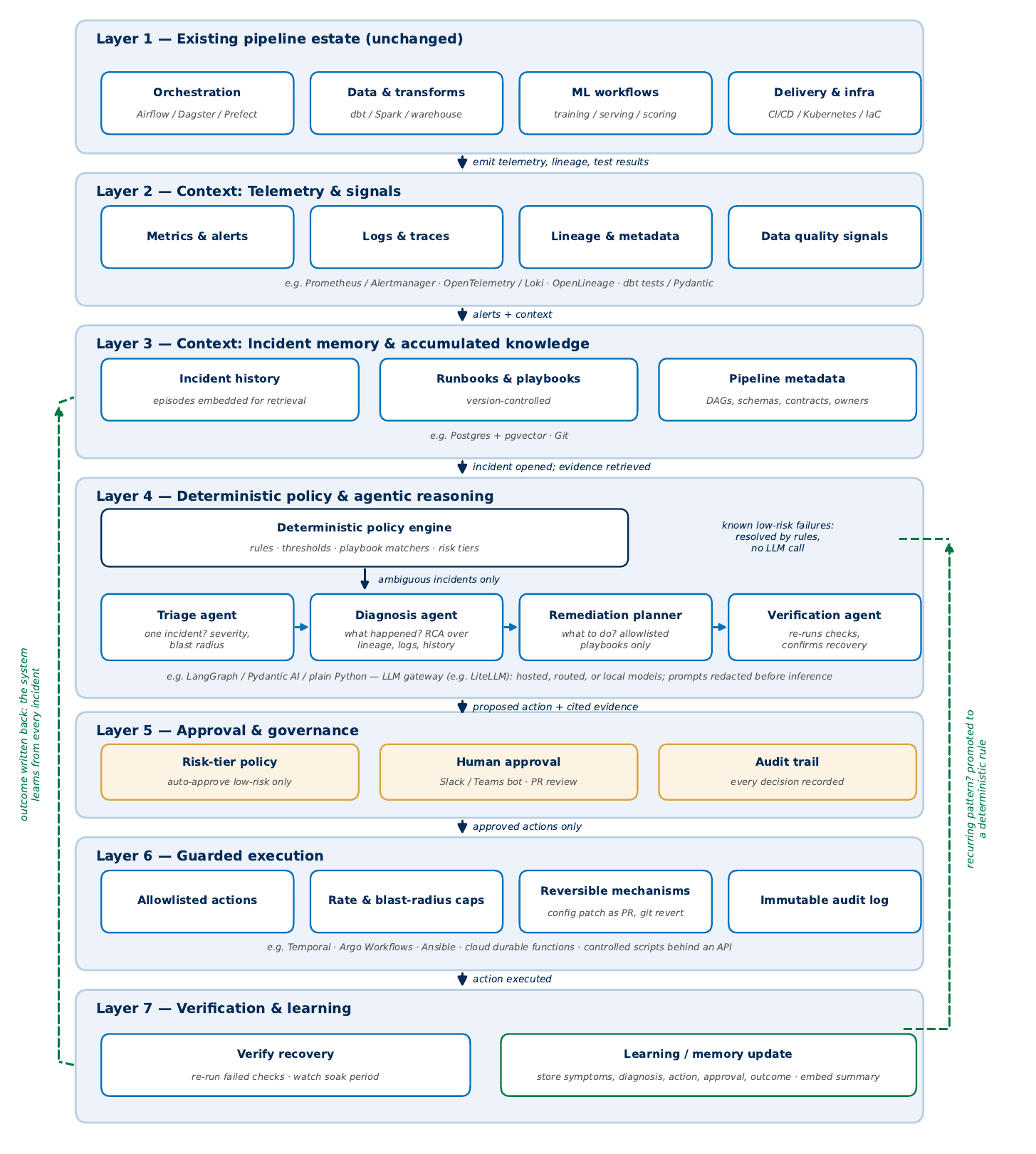}
  \caption{The proposed reference architecture as seven logical layers. The existing pipeline estate (top) is instrumented through open standards; alerts flow through the incident memory layer into a reasoning layer where a deterministic policy engine handles known failures and four LLM agents handle ambiguous ones; proposed actions pass an approval and governance layer (shown in amber, marking where a human is in the loop) before reaching a guarded execution layer; and verification outcomes are written back into incident memory (left, dashed green arrow), closing the learning loop. A second feedback path (right, dashed green arrow) promotes recurring diagnosis-and-remediation patterns from Layer 7 into new rules in the Layer 4 deterministic policy engine. Tools named in small print are illustrative examples, not required components.}
  \label{fig:architecture}
\end{figure}

Figure~\ref{fig:architecture} shows the seven logical layers of the architecture. Layers 2 and 3 are explicitly labelled \emph{Context} because they provide the live evidence and accumulated operational knowledge that the reasoning layer consults; they do not themselves decide or execute remediations. Each layer is defined by a responsibility and an interface; the example tools noted in the figure and below are replaceable implementations of those responsibilities, not requirements.

\textbf{Layer 1: Existing pipeline estate.} The systems being healed---orchestrators (e.g., Airflow, Dagster, Prefect), transformation and compute layers (e.g., dbt, Spark, warehouses), ML workflows (experiment tracking, model serving, batch scoring), and delivery infrastructure (CI/CD, Kubernetes, infrastructure-as-code). The architecture requires no replacement of any of these; it only requires that they emit telemetry.

\textbf{Layer 2: Context: Telemetry \& signals.} Four signal families, each an interface with mature open implementations: metrics and alerting (e.g., an existing monitoring system, or Prometheus with Alertmanager~\cite{prometheus2025}), logs and traces (e.g., OpenTelemetry collectors with a store such as Loki), lineage and catalog metadata (e.g., OpenLineage events into Marquez or OpenMetadata~\cite{openlineage2025}), and data quality signals---lightweight data quality checks, dbt tests, or optional tools such as Pydantic~\cite{dbttests2026}. Model-quality signals---drift statistics, evaluation metrics---are exported as ordinary metrics so that no separate ML monitoring stack is needed at the start.

\textbf{Layer 3: Context: Incident memory and accumulated knowledge.} A deliberately boring core: a relational database (e.g., Postgres) holding incidents, their timelines, and their resolutions, with text fields embedded into a vector index for similarity retrieval; runbooks and remediation playbooks maintained as Markdown in Git; and a pipeline metadata registry (DAG definitions, schemas, data contracts, ownership) assembled from the lineage layer. Over time, resolved incidents accumulate evidence about recurring failure signatures, effective actions, and appropriate approvals. Retrieval makes that accumulated knowledge available during diagnosis while still requiring each proposed action to be checked against current evidence and policy. This layer is what converts one-off firefighting into organizational memory (P5).

\textbf{Layer 4: Deterministic policy and agentic reasoning.} The reasoning layer has two tiers. Incidents first pass through a \emph{deterministic policy engine}: version-controlled rules, thresholds, playbook matchers, and risk classifications that recognize known failure signatures and route them directly to a proven remediation without any LLM call (P3). Only incidents the rules cannot resolve---ambiguous evidence, novel symptoms, cross-system failures---are escalated to four cooperating LLM agents, each an LLM invocation with task-specific tools and prompts following the reason-and-act pattern~\cite{yao2023}. The four agents have deliberately separate responsibilities:
\begin{itemize}[leftmargin=1.5em]
  \item The \emph{triage agent} decides whether related alerts belong to a single incident, classifies the failure against the taxonomy of Section~\ref{sec:failures}, estimates blast radius from lineage (which downstream assets and consumers are affected), and assigns severity.
  \item The \emph{diagnosis agent} investigates \emph{what happened}: it retrieves similar past incidents, walks the lineage graph upstream from the failing asset, queries logs and metrics around the failure window, and produces a root-cause hypothesis with cited evidence. Grounding in retrieved history and live telemetry---rather than free generation---is the key defense against confabulated diagnoses~\cite{ahmed2023,chen2024}.
  \item The \emph{remediation planner} decides \emph{what should be done}: it maps the diagnosis to a candidate action \emph{from the allowlisted playbook library only}. Diagnosis and planning are intentionally distinct responsibilities---one answers ``what happened?'', the other ``what should we do?''---so each can be audited, evaluated, and improved separately. If no playbook applies, the planner's only permitted output is an escalation with a well-structured summary---an outcome that is itself valuable, since it hands the on-call engineer a completed investigation.
  \item The \emph{verification agent} re-runs the checks that detected the failure, watches recovery metrics for a configurable soak period, and either closes the incident or escalates with full context.
\end{itemize}
A single tool-using agent could perform all four roles, and for a small deployment that may be the simpler choice. The four-way split trades some orchestration complexity, latency, and inference cost for a benefit that matters more in a guarded, auditable loop: each responsibility can be scoped, evaluated, and improved independently, and one agent's failure mode---a bad diagnosis, say---is isolated rather than compounding inside a single large tool-using loop. As a reference architecture, this separation is what we recommend, not a requirement.

The control plane connecting these agents is a small state graph, and any agent orchestration approach can implement it: LangGraph, Pydantic AI, Agno, Semantic Kernel, AutoGen, plain Python services, or a custom state machine~\cite{langgraph2026,pydanticai2026}. Routing between agents and their tools is driven by the triage agent's classification: a data-quality incident is routed to lineage and schema-diff evidence, an infrastructure incident to logs and metrics, and so on. This routing is implemented as conditional edges in the state graph rather than as a separate router agent, since the classification needed to route already exists as the triage agent's output. All agents call models through an \emph{LLM gateway} (e.g., LiteLLM or a thin internal proxy~\cite{litellm2026}), behind which hosted APIs, model routing providers, and locally deployed open-weight models are interchangeable depending on cost, privacy, latency, and data-residency requirements (P2). The gateway also enables policy-based routing: high-volume triage tasks may use cheaper hosted models, sensitive incidents may be restricted to locally hosted models, and complex diagnosis tasks may use stronger frontier models when allowed by data-governance policy. Before incident context is passed to any model, the gateway applies data minimization and redaction: secrets, tokens, personal data, customer identifiers, and unnecessary raw logs are removed or masked, and agents receive only the evidence required for diagnosis rather than unrestricted access to the full production environment.

\textbf{Layer 5: Approval and governance.} Between planning and execution sits a policy gate implemented where engineers already work: a Slack/Teams bot for operational approvals and pull-request review for code-level changes. Listing~\ref{lst:policy} shows the shape of a risk-tier policy. Low-risk, reversible actions (retry a task, refresh a materialized view) may be auto-approved; medium-risk actions require one approval; high-risk actions (schema migrations, production deployments, anything touching data deletion) always require a human and are often better left as recommendation-only. Every decision---automatic or human---is recorded in the audit trail.

\noindent\begin{minipage}{\linewidth}
\begin{lstlisting}[language={},caption={Example risk-tier policy for remediation actions (YAML).},label={lst:policy}]
actions:
  retry_task:
    risk: low
    auto_approve: true
    max_per_day: 5          # per pipeline
  backfill_partition:
    risk: medium
    auto_approve: false     # one approver in Slack
    requires: [diagnosis_confidence >= 0.8]
  rollback_deployment:
    risk: medium
    auto_approve: false
    mechanism: git_revert_pr
  alter_schema:
    risk: high
    auto_approve: never     # recommendation only;
                            # human executes via PR
\end{lstlisting}
\end{minipage}

\textbf{Layer 6: Guarded execution.} Approved actions execute through a guarded workflow execution layer rather than by giving agents direct shell access. Examples include Temporal, Argo Workflows, Ansible, cloud-native durable functions, or controlled internal scripts behind an API~\cite{temporal2025,argo2026}. The execution layer enforces the allowlist, applies rate limits and blast-radius caps (e.g., at most $N$ automated actions per pipeline per day), executes changes preferentially through reviewable mechanisms (a config patch becomes a pull request; a rollback becomes a git revert plus redeploy), and writes an immutable audit record of every action, its initiator, and its approval trail.

\textbf{Layer 7: Verification and learning.} The loop does not end when an action executes. The verification agent confirms recovery, and a \emph{learning/memory update} step then writes the complete episode back into the incident memory layer: symptoms, diagnosis, selected action, approval decision, remediation outcome, verification result, and final resolution. Each episode is tagged with its verification provenance---auto-verified after the soak period, or human-confirmed---since a coincidental recovery should not be retrieved with the same confidence as a precedent a human has reviewed. Incident summaries are embedded for similarity search so that future diagnoses retrieve and reuse them. This step is what makes the memory of Layer 3 compound over time (P5): every resolved incident makes the next diagnosis faster and better grounded. A pattern that recurs across several episodes with the same diagnosis and remediation is itself a signal worth acting on: it is a natural candidate for promotion into the deterministic policy engine of Layer 4, giving the architecture an explicit path from agentic diagnosis toward deterministic automation as failure modes become well understood.

\subsection{Example implementation options and indicative cost}
\label{sec:cost}

Table~\ref{tab:tools} lists example implementation options for each architectural concern. The tools listed are illustrative rather than prescriptive: the architecture is defined by interfaces and responsibilities, not by a fixed technology stack, and a practical implementation should select the smallest viable set of tools that fits the organization's existing environment. In most rows the minimal implementation is a tool the team already runs. Appendix~\ref{app:tools} compiles a directory of all tools and platforms mentioned in this paper, with official links.

\begin{table}[H]
  \centering
  \caption{Example implementation options for the reference architecture. Tools are illustrative, not prescriptive; start from the minimal column and substitute or extend only where the environment requires it.}
  \label{tab:tools}
  \footnotesize
  \begin{tabular}{@{}p{2.7cm}p{3.6cm}p{4.2cm}p{3.6cm}@{}}
    \toprule
    \textbf{Architectural concern} & \textbf{Minimal implementation} & \textbf{Possible substitutes or extensions} & \textbf{Notes} \\
    \midrule
    Metrics \& alerts & Existing monitoring system, or Prometheus + Alertmanager & Grafana, Datadog, CloudWatch, Azure Monitor & Reuse whatever already pages the team \\
    \lightrule
    Logs \& traces & Existing log store, or OpenTelemetry + Loki & ELK, Grafana Tempo, Datadog Logs & Structured logs around failure windows are what agents consume \\
    \lightrule
    Lineage \& metadata & Orchestrator metadata and pipeline definitions & OpenLineage, OpenMetadata, Marquez & A dedicated lineage platform is optional at the start \\
    \lightrule
    Data quality checks & dbt tests or lightweight custom checks & Pydantic & Results exported as metrics; keep validation lightweight \\
    \lightrule
    Incident \& memory store & Postgres + pgvector & SQLite, Qdrant, OpenSearch & Boring by design; embedded summaries enable retrieval \\
    \lightrule
    Deterministic policy layer & Version-controlled rules and risk tiers (YAML + a small rules service) & Open Policy Agent, custom rules engine & Handles known failures before any LLM call \\
    \lightrule
    LLM gateway & LiteLLM or a thin internal proxy & Direct provider SDKs, internal routing service & Hosted APIs, routing providers, and local open-weight models interchangeable \\
    \lightrule
    Agent orchestration & LangGraph or plain Python services & Pydantic AI, Agno, Semantic Kernel, AutoGen, custom state machine & The control plane is a small state graph; any framework---or none---can express it \\
    \lightrule
    Approval workflow & Slack or Teams bot & GitHub/GitLab pull-request review & Approvals live where engineers already work \\
    \lightrule
    Guarded actuation & Existing scripts behind a controlled API & Temporal, Argo Workflows, Ansible & Durable, audited, rate-limited execution \\
    \bottomrule
  \end{tabular}
\end{table}

The steady-state cost has three components: (i) infrastructure for the monitoring and agent services---a few small VMs or a modest Kubernetes namespace; (ii) LLM inference, which is dominated by diagnosis calls and is bounded by incident volume rather than estate size (a team handling tens of incidents per week should expect LLM spend in the tens-to-hundreds of dollars per month range, depending on model choice)---and which the deterministic policy layer further reduces by resolving known failures without any model call; and (iii) engineering time, the honest headline cost, discussed in Section~\ref{sec:discussion}. Crucially, cost scales with \emph{incidents}, not with hosts, tables, or seats---the inverse of the commercial pricing models surveyed in Section~\ref{sec:landscape}.

\begin{figure}[H]
  \centering
  \includegraphics[width=0.75\textwidth]{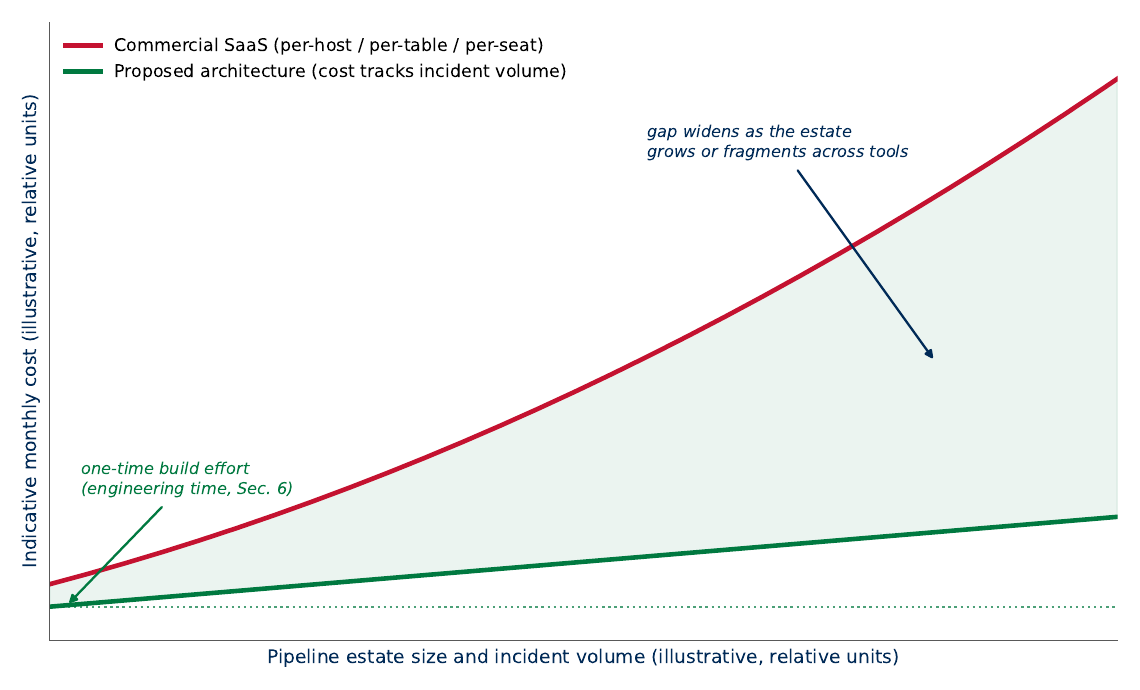}
  \caption{Illustrative cost scaling: commercial per-host/per-table/per-seat pricing (as documented for Dynatrace, Datadog, Monte Carlo, and PagerDuty in Section~\ref{sec:landscape}) grows with estate size, while the proposed architecture's cost grows with incident volume instead, after a one-time build effort. This chart tracks infrastructure and inference cost only; the ongoing engineering effort to operate and maintain the stack is a separate, real cost discussed in Section~\ref{sec:discussion}. Curves are qualitative, intended to illustrate the pricing-model inversion argued for in this section, not measured data.}
  \label{fig:cost-scaling}
\end{figure}

\section{The Self-Healing Lifecycle in Practice}
\label{sec:lifecycle}

\begin{figure}[H]
  \centering
  \includegraphics[width=0.82\textwidth]{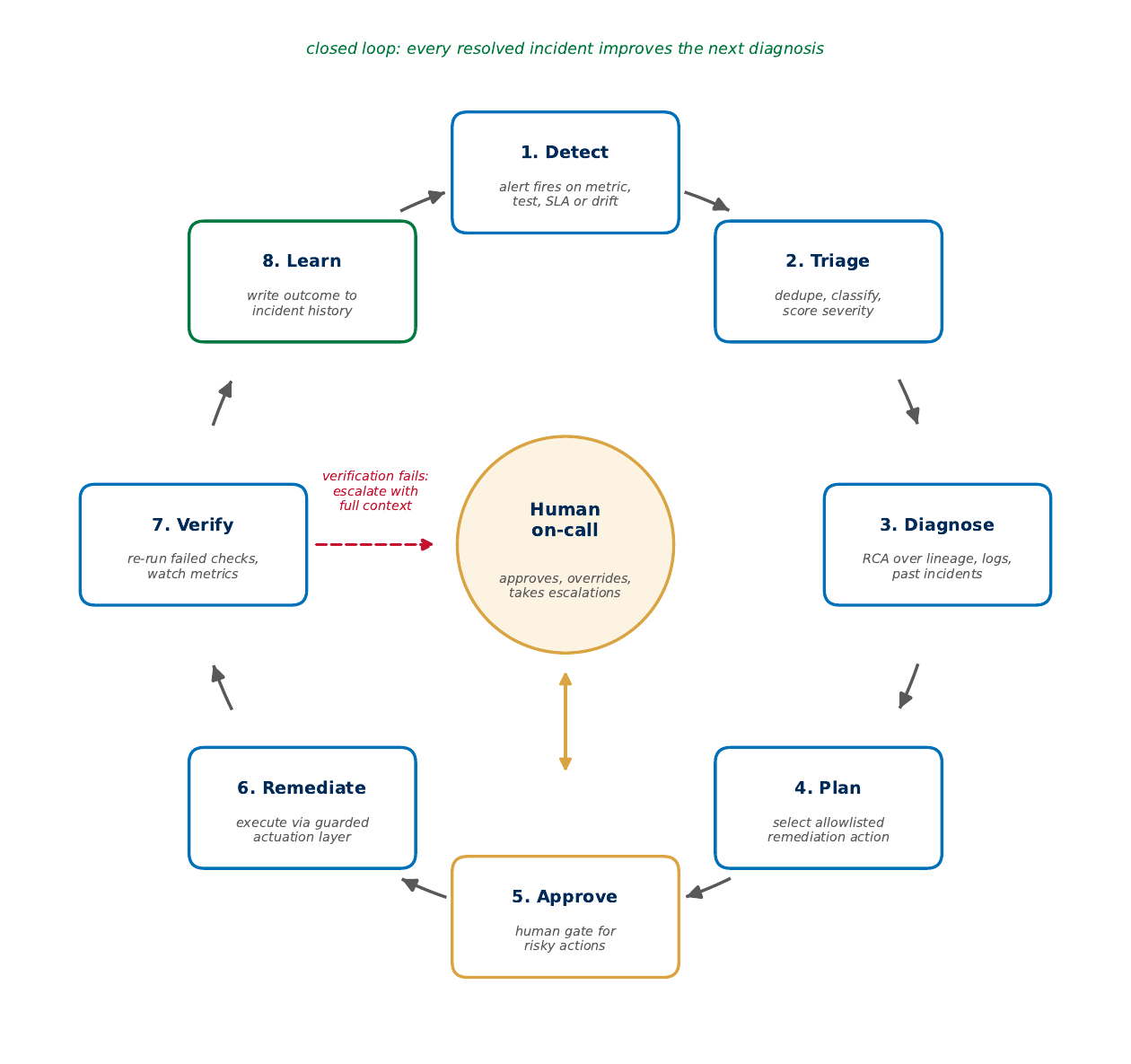}
  \caption{The self-healing incident lifecycle. Stages 1--4 and 6--8 are supported by a combination of deterministic checks, workflow logic, and AI agents; stage 5 remains a policy-controlled approval gate, shown in amber to mark the human-in-the-loop checkpoint. Low-risk reversible actions may be auto-approved, while high-risk or business-sensitive actions require human review. The human on-call engineer approves risky actions and receives escalations with full investigative context when verification fails.}
  \label{fig:lifecycle}
\end{figure}

Figure~\ref{fig:lifecycle} shows the eight-stage self-healing lifecycle, with the human on-call engineer at the center: consulted at the approval gate, and handed a completed investigation whenever verification fails or no playbook applies.

Three worked examples illustrate the loop across the pipeline domains. To keep them technology-flexible we use general terms---an orchestrator task, a schema validation check---naming specific tools only where the detail matters. 

\textbf{Schema drift (data engineering).} An upstream team renames \texttt{customer\_id} to \texttt{cust\_id} in a source export. A lightweight schema validation check fails at ingestion (\emph{detect}); the triage agent classifies the incident as a schema change and, from lineage, flags fourteen downstream models and two dashboards as at risk (\emph{triage}). The diagnosis agent compares the failing schema against the last successful run's schema, identifies the rename, and retrieves a similar incident from four months earlier (\emph{diagnose}). The planner selects the \texttt{column\_rename\_patch} playbook, which generates a pull request updating the staging model's column mapping plus a data contract violation notice to the upstream owner (\emph{plan}). Because the action is a reviewable PR, the policy tier is medium: one engineer approves (\emph{approve}); CI runs, the patch merges, and the actuation layer triggers a backfill of the affected partition (\emph{remediate}). The verification agent re-runs the failed validation suite and confirms downstream freshness (\emph{verify}), then writes the episode---including the upstream notification---to the incident store (\emph{learn}).

\textbf{Stuck orchestration (infrastructure).} An orchestrator sensor task (here, an Airflow sensor) deadlocks after a spot-instance preemption kills its worker mid-poll. The alerting layer fires on task-duration and worker-loss metrics; the triage agent merges both alerts into one incident. Diagnosis correlates the worker termination event in the logs with the stuck sensor and finds three prior identical incidents, all resolved by clearing and re-running the task. The planner selects \texttt{retry\_task}---low-risk, auto-approved, under its daily rate cap---and the actuation layer clears the task via the orchestrator API. Verification watches the DAG complete. Total human involvement: a Slack notification read after the fact. Notably, a failure signature this well-established is exactly what the deterministic policy layer is for: after enough recurrences, a rule can route it straight to the playbook with no LLM reasoning at all.

\textbf{Model performance decay (MLOps).} A weekly churn model's exported performance metric (here, AUC) drifts below its alert threshold. Triage classifies a model workflow issue with low urgency but high business impact. Diagnosis inspects feature drift statistics, finds that a promotional campaign shifted the distribution of two behavioral features, and notes---from the incident store---that retraining resolved a similar episode previously. The planner proposes the \texttt{trigger\_retraining} playbook, but the policy requires human approval for model releases: the on-call ML engineer reviews the drift evidence, approves retraining into a staging slot, and the verification agent compares champion/challenger metrics before the human makes the final promotion decision. The loop here is deliberately only half-automated: the contribution is that the engineer received a completed investigation and a one-click decision rather than a bare alert~\cite{breck2017}.

Across all three examples the pattern is consistent. The goal here is not maximal autonomy, but \emph{controlled autonomy}: agents compress investigation and execution mechanics, while irreversible, high-risk, or business-sensitive decisions remain under human control. In our view this division---rather than maximal autonomy---is the correct target for the current generation of LLM agents.

These examples also show that self-healing is not a single action, but a controlled lifecycle. The architecture does not simply detect an error and execute a fix. It gathers evidence, identifies the likely cause, selects from approved remediation options, applies policy-based approval, verifies recovery, and records the outcome. This lifecycle is what makes the system safer, auditable, and reusable across different pipeline environments---and it is the core of the paper's contribution.

\section{Discussion}
\label{sec:discussion}

\textbf{When to buy instead.} The proposed architecture is not a categorical argument against commercial platforms. A team already standardized on Databricks obtains much of this capability by turning it on~\cite{databricks2025}; an enterprise with a mature ServiceNow practice may rationally extend it~\cite{servicenow2025}; and a team with no operations bandwidth at all may be better served by a managed product than by any reference architecture. The build path is most attractive when the estate is heterogeneous, when per-host/per-table pricing is prohibitive relative to team size, when data residency rules out SaaS telemetry, or when vendor exit costs are a strategic concern. A useful heuristic is that the build path becomes attractive when the annual cost of commercial coverage exceeds the engineering effort required to maintain a minimal implementation. This estimate is context-dependent and should be validated against the organization's incident volume, existing tooling, and available platform engineering capacity.

\textbf{Engineering cost is the real price.} The components in Table~\ref{tab:tools} are free to license, not free to run. Standing up the telemetry layer, writing the first dozen playbooks, and tuning alert thresholds is genuine work, and skipping it does not make the work disappear---it relocates it into every future incident. Principle P6 mitigates the risk: each layer pays for itself independently, and the highest-value early deliverable is usually not remediation at all but the diagnosis agent, which converts every alert into a completed investigation. The practical starting point should be the smallest useful loop: ingest alerts from existing tools, create an incident record, generate a diagnosis summary, route it to the responsible engineer, and store the outcome. Automated remediation, lineage integration, and advanced policy enforcement can be added only after the diagnosis loop proves useful.

\textbf{Risks of LLM-driven operations.} Three failure modes deserve explicit treatment. \emph{Confabulated diagnoses}: an LLM will produce a fluent root-cause narrative whether or not the evidence supports one; grounding in retrieved incidents and live telemetry, requiring cited evidence, and reporting calibrated confidence are necessary mitigations~\cite{ahmed2023,chen2024}. \emph{Automation bias}: approvals degrade into rubber stamps when the agent is usually right; rotating human review of auto-approved actions, and periodically sampling agent diagnoses for audit, keep the human check meaningful. \emph{Compounding actions}: a wrong remediation can trigger further alerts and further remediations; the rate caps, blast-radius limits, and preference for reversible mechanisms in the execution layer exist precisely to bound this failure mode. Where possible, known failure patterns should be handled by deterministic rules and playbooks before invoking LLM reasoning at all; this reduces cost, improves predictability, and limits the role of LLMs to ambiguous incidents that require contextual reasoning across logs, lineage, runbooks, and incident history. More broadly, an agent with actuation credentials is itself new attack surface, and the audit trail must be treated as a security control, not an afterthought.

\textbf{Data privacy and provider governance.} A further concern arises when LLMs are used for operations. Incident context may include logs, traces, customer identifiers, infrastructure names, source code snippets, database schemas, or commercially sensitive information. Sending this context to an external model provider may be unacceptable in regulated or security-sensitive environments---not only as a technical risk but as a governance and trust concern, since providers differ in whether prompts are retained, whether submitted data is used for model training, and what enterprise controls, data-residency options, and audit logs are available. The architecture therefore treats model access as a policy-controlled gateway rather than a direct dependency on a single provider. Sensitive incidents can be routed to locally hosted or enterprise-approved models; prompts are redacted and minimized before inference (Section~\ref{sec:architecture}); and provider settings on data retention, model training, and logging must be verified against current official documentation before deployment.

\textbf{Limitations.} This paper is prescriptive, not empirical: the architecture distills patterns from the surveyed platforms and published RCA research rather than reporting a longitudinal deployment, and the cost claims of Section~\ref{sec:cost} are indicative rather than measured. The platform comparison reflects vendor capabilities as publicly described in mid-2026 and will age quickly. The worked examples assume failure classes with recognizable precedents; genuinely novel incidents will always escalate to humans, and a system evaluated only on routine incidents will overstate its coverage. Finally, regulated environments may require stricter approval, explainability, and change-control regimes than the policy model presented here; the governance-first posture of platforms like ServiceNow's AI Control Tower~\cite{servicenow2025} indicates the direction such extensions must take. Future work should instrument real deployments of this architecture and report incident-level outcomes---time to diagnosis, time to recovery, escalation rates, and remediation precision---against a pre-adoption baseline, following the outcome-oriented measurement tradition of the DevOps literature~\cite{forsgren2018}.

\section{Conclusion}
\label{sec:conclusion}

Pipeline operations sit at an awkward point in the market: the platforms that heal most autonomously demand the deepest ecosystem commitment and the largest budgets, while the teams that suffer most from pipeline failures---small, heterogeneous, cost-constrained---are precisely those least able to adopt them. This concern is sharpest for small data teams, organizations in emerging markets, and any environment where tooling is heterogeneous and budgets are tight; for them, affordability and vendor independence are not preferences but preconditions. This paper argued that the gap is architectural rather than technological. By architectural gap, we mean that the missing piece is not a single new algorithm or a completely new class of infrastructure tool; rather, it is a coherent design that connects telemetry, metadata, incident memory, deterministic policy, agentic diagnosis, approval workflows, guarded execution, verification, and learning into one controlled recovery loop. The mechanisms underneath commercial ZeroOps offerings---telemetry, metadata, inference, approval, actuation---are convergent and individually available as mature open-source components, and LLM agents can provide a flexible reasoning layer over telemetry, lineage, runbooks, and incident history, complementing rather than replacing deterministic rules and classical correlation methods. The reference architecture presented here assembles these components into a guarded self-healing loop: deterministic checks and agents triage incidents, diagnose against lineage and incident history, plan only allowlisted actions, verify recovery, and write outcomes back into organizational memory; humans approve what is risky and receive completed investigations for what is not. Its cost scales with incidents rather than estate size, its components are individually replaceable, and its value compounds through the incident memory it accumulates. For data engineering, MLOps, and software delivery teams alike, this architecture offers a guarded, memory-centered, vendor-agnostic pattern for building pipelines that increasingly diagnose, recover, and learn from failures with less manual effort.

\bibliographystyle{unsrt}
\bibliography{references}

\appendix
\setcounter{table}{0}
\renewcommand{\thetable}{\thesection\arabic{table}}

\section{Proof-of-Concept Implementation}
\label{app:poc}

A minimal proof of concept accompanies this paper: a single Python file running the full detect--triage--diagnose--plan--approve--remediate--verify--learn loop end to end.

\begin{itemize}[leftmargin=1.5em]
  \item \textbf{Repository:} \url{https://github.com/soloshun/agentic-recovery-and-incident-response}
\end{itemize}

\section{Tools and Platforms Mentioned in the Paper}
\label{app:tools}

This appendix lists the tools and platforms mentioned in the paper. The list is intended as a practical directory, not an endorsement or a required implementation stack. The proposed architecture is tool-agnostic; these tools are examples of components that could satisfy different architectural responsibilities. Links point to official product pages or documentation.

\begin{table}[H]
  \centering
  \caption{Off-the-shelf operations platforms surveyed in Section~\ref{sec:landscape}, and adjacent AI-assisted tools discussed in the text. All are commercial products (some with free tiers).}
  \label{tab:app-platforms}
  \footnotesize
  \begin{tabular}{@{}p{3.4cm}p{4.2cm}p{6.6cm}@{}}
    \toprule
    \textbf{Tool / platform} & \textbf{Role in the paper} & \textbf{Official link} \\
    \midrule
    Databricks Genie ZeroOps & Surveyed platform (Sec.~\ref{sec:landscape}) & \scriptsize\url{https://www.databricks.com/blog/introducing-genie-zeroops} \\
    Acceldata ADM & Surveyed platform & \scriptsize\url{https://www.acceldata.io/adm} \\
    Dynatrace Davis AI & Surveyed platform & \scriptsize\url{https://docs.dynatrace.com/docs/discover-dynatrace/platform/davis-ai} \\
    Datadog Bits AI & Surveyed platform & \scriptsize\url{https://www.datadoghq.com/product/platform/bits-ai/} \\
    Monte Carlo & Surveyed platform & \scriptsize\url{https://www.montecarlodata.com/product/data-observability-platform/} \\
    IBM Databand & Surveyed platform & \scriptsize\url{https://www.ibm.com/products/databand} \\
    PagerDuty SRE Agent & Surveyed platform & \scriptsize\url{https://www.pagerduty.com/platform/ai-agents/sre/} \\
    ServiceNow Predictive AIOps & Surveyed platform & \scriptsize\url{https://www.servicenow.com/products/predictive-aiops.html} \\
    \lightrule
    Sentry & Adjacent: AI-assisted application error monitoring (Sec.~\ref{sec:background}) & \scriptsize\url{https://sentry.io/} \\
    Prefect Marvin & Adjacent: agentic AI framework (Sec.~\ref{sec:landscape}) & \scriptsize\url{https://github.com/PrefectHQ/marvin} \\
    LangSmith Engine & Adjacent: LLM application trace observability and failure diagnosis & \scriptsize\url{https://docs.langchain.com/langsmith/engine} \\
    \bottomrule
  \end{tabular}
\end{table}

\begin{table}[H]
  \centering
  \caption{Open-source and low-cost components usable in the reference architecture. All are open source unless noted.}
  \label{tab:app-oss}
  \footnotesize
  \begin{tabular}{@{}p{3.0cm}p{4.6cm}p{6.6cm}@{}}
    \toprule
    \textbf{Tool} & \textbf{Architectural concern} & \textbf{Official link} \\
    \midrule
    Prometheus & Metrics \& alerting & \scriptsize\url{https://prometheus.io/} \\
    Alertmanager & Metrics \& alerting & \scriptsize\url{https://prometheus.io/docs/alerting/latest/alertmanager/} \\
    Grafana & Dashboards \& visualization & \scriptsize\url{https://grafana.com/} \\
    OpenTelemetry & Logs \& traces & \scriptsize\url{https://opentelemetry.io/} \\
    Grafana Loki & Log store & \scriptsize\url{https://grafana.com/oss/loki/} \\
    Grafana Tempo & Trace store & \scriptsize\url{https://grafana.com/oss/tempo/} \\
    OpenLineage & Lineage standard & \scriptsize\url{https://openlineage.io/} \\
    Marquez & Lineage store & \scriptsize\url{https://marquezproject.ai/} \\
    OpenMetadata & Metadata \& catalog & \scriptsize\url{https://open-metadata.org/} \\
    dbt tests & Data quality checks & \scriptsize\url{https://docs.getdbt.com/docs/build/data-tests} \\
    Pydantic & Schema validation & \scriptsize\url{https://docs.pydantic.dev/} \\
    PostgreSQL & Incident \& memory store & \scriptsize\url{https://www.postgresql.org/} \\
    pgvector & Similarity retrieval & \scriptsize\url{https://github.com/pgvector/pgvector} \\
    Git & Runbooks, playbooks, audit & \scriptsize\url{https://git-scm.com/} \\
    Open Policy Agent & Deterministic policy layer & \scriptsize\url{https://www.openpolicyagent.org/} \\
    LangGraph & Agent orchestration & \scriptsize\url{https://www.langchain.com/langgraph} \\
    LangChain & Agent framework & \scriptsize\url{https://www.langchain.com/} \\
    Pydantic AI & Agent framework & \scriptsize\url{https://ai.pydantic.dev/} \\
    Agno & Agent framework & \scriptsize\url{https://github.com/agno-agi/agno} \\
    Semantic Kernel & Agent framework & \scriptsize\url{https://learn.microsoft.com/en-us/semantic-kernel/} \\
    AutoGen & Agent framework & \scriptsize\url{https://microsoft.github.io/autogen/} \\
    Temporal & Guarded execution & \scriptsize\url{https://temporal.io/} \\
    Argo Workflows & Guarded execution & \scriptsize\url{https://argoproj.github.io/workflows/} \\
    Ansible & Guarded execution & \scriptsize\url{https://www.ansible.com/} \\
    Slack & Approval workflow (commercial) & \scriptsize\url{https://slack.com/} \\
    Microsoft Teams & Approval workflow (commercial) & \scriptsize\url{https://www.microsoft.com/en-us/microsoft-teams/group-chat-software} \\
    GitHub & PR review, code hosting (commercial) & \scriptsize\url{https://github.com/} \\
    GitLab & PR review, code hosting (commercial) & \scriptsize\url{https://gitlab.com/} \\
    \bottomrule
  \end{tabular}
\end{table}

\begin{table}[H]
  \centering
  \caption{LLM providers, routing, and model serving. Model-provider choice is implementation-specific: teams should select providers based on cost, latency, model quality, data privacy, data residency, enterprise controls, and organizational compliance requirements. This list is not an endorsement and does not imply that all providers are suitable for regulated operations.}
  \label{tab:app-llm}
  \footnotesize
  \begin{tabular}{@{}p{3.0cm}p{4.6cm}p{6.6cm}@{}}
    \toprule
    \textbf{Tool / provider} & \textbf{Category} & \textbf{Official link} \\
    \midrule
    OpenAI & Hosted model provider & \scriptsize\url{https://openai.com/} \\
    Anthropic & Hosted model provider & \scriptsize\url{https://www.anthropic.com/} \\
    Google Gemini & Hosted model provider & \scriptsize\url{https://ai.google.dev/} \\
    Groq & Hosted inference provider & \scriptsize\url{https://groq.com/} \\
    Together AI & Hosted inference provider & \scriptsize\url{https://www.together.ai/} \\
    Fireworks AI & Hosted inference provider & \scriptsize\url{https://fireworks.ai/} \\
    OpenRouter & Model routing provider & \scriptsize\url{https://openrouter.ai/} \\
    LiteLLM & LLM gateway (open source) & \scriptsize\url{https://github.com/BerriAI/litellm} \\
    vLLM & Self-hosted model serving (open source) & \scriptsize\url{https://docs.vllm.ai/} \\
    Ollama & Local model serving (open source) & \scriptsize\url{https://ollama.com/} \\
    llama.cpp & Local inference (open source) & \scriptsize\url{https://github.com/ggml-org/llama.cpp} \\
    \bottomrule
  \end{tabular}
\end{table}

\section*{Declaration of Generative AI Use}

The authors used Claude (Anthropic) in the preparation of this work.
For code development, the tool was used to accelerate syntax generation, debugging, and boilerplate code structuring, including the scripts that render the figures; the foundational logic and designs were conceptualized by the authors.
For writing, the tool was used to improve language, grammar, and readability.
After using this tool, the authors thoroughly reviewed, edited, and verified all resulting text and code, and assume full responsibility for the accuracy and integrity of the final contents of this paper.

\end{document}